# Superconductivity From Repulsive Electronic Correlations on Alternant Cuprate and Iron-based Lattices

Lawrence J. Dunne [a, b, 1], Erkki J. Brändas [c]

[a] Department of Materials, Imperial College London, London SW7 2AZ, UK
[b] Department of Engineering Systems, London South Bank University, London SE1 0AA, UK
[c] Department of Chemistry, Ångström Laboratory, University of Uppsala, Box 518, Uppsala, S-75120, Sweden

**Abstract**

A key question in the theory of high-temperature superconductivity is whether Off-diagonal Long-Range Order (ODLRO) can be induced wholly or in large part by repulsive electronic correlations. Electron pairs on Cuprate and the iron-based pnictide and chalcogenide alternant lattices may interact with a strong short-range Coulomb repulsion and much weaker longer range attractive tail. Here we show that such interacting electrons can cooperate to produce a superconducting state in which time-reversed electron pairs effectively avoid the repulsive part but reside predominantly in the attractive region of the potential. The alternant lattice structure is a key feature of such a stabilization mechanism leading to the occurrence of high temperature superconductivity with $d_{x^2-y^2}$ or sign alternating s-wave or $s \pm$ condensate symmetries.

-------------------------------------------------------------------------------------------------------

1- Visiting Professor

A key question currently in the theory of high-temperature superconductivity is whether Off-diagonal Long-Range Order (ODLRO) can be induced wholly or in large part by repulsive electronic correlations. Although it is now some years since the discovery of high Tc superconducting cuprates[1] and more recently the iron-based superconductors[2,3], the nature of the superconducting condensate in such materials continues to be intensely discussed[4,5]. Following phase sensitive detection [6,7,8,9] and ARPES[10,11,12] and a range of other experimental techniques[13,14] there is now a widely held view that the pair condensate wavefunctions have $d_{x^2-y^2}$ and s± symmetries respectively in superconducting cuprates and iron-based materials with possible nodal or nodeless features in the latter [15,13,16].

These classes of superconducting materials have alternant structures in which every unit cell may be given a positive or negative sign which can be arranged such that each cell is surrounded with cells of opposite signature. Here we show that such an alternant structural feature of cuprates and iron-based superconductors allows a remarkable energetic stabilization mechanism when electrons interact via a screened but strong short-range repulsion and a much weaker long-range attractive tail as is possible in variety of scenarios such as spin-fluctuations or Friedel oscillations or other mechanisms but where the Hamiltonian matrix has repulsive off-diagonal elements. Real-space pairing occurs leading to the occurrence of superconductivity with $d_{x^2-y^2}$ and sign alternating s-wave or s± condensate wavefunctions. Electrons can cooperate to produce a superconducting state in which electron pairs effectively avoid the repulsive part of the potential but reside predominantly in the attractive region. The alternant lattice structure is a key feature of such a stabilization mechanism. We do not identify the cause of the attractive region of the electron-electron interaction but show how real-space electronic pairing on a lattice can exploit such a potential, for which there are numerous candidates.

In conventional BCS superconductors[17] with an attractive phonon induced electron-electron interaction it is commonplace to discuss the instability of the Fermi sea to superconducting electron pair formation in terms of an n-fold stabilization problem[17,18]. Thus, if n quasi-degenerate states at energy $U$ above the Fermi level are coupled by an attractive off-diagonal matrix element $-V$, then the Hamiltonian matrix has (n-1) eigenvalues at energy $U$ and one at energy $U$-nV which splits off from the others and crosses over into a re-built ground state.

Following Yang[19] the connection with superconductivity can be made from an electron pair population analysis using the second order reduced electronic density matrix[20,21,22,23,24] $\rho_2(\mathbf{x_1},\mathbf{x_2};\mathbf{x_1}',\mathbf{x_2}')$ which demonstrates the existence of Off-diagonal Long-range Order (ODLRO)[19] in the ground state wavefunction. Thus, for a 2$M$- electron wavefunction $\Psi(\mathbf{x}_1,\mathbf{x}_2,\ldots\mathbf{x}_{2M})$ the second order reduced electronic density matrix is given by

$$\rho_2(\mathbf{x_1},\mathbf{x_2};\mathbf{x_1}',\mathbf{x_2}') = 2M(2M-1)\int \Psi(\mathbf{x}_1,\mathbf{x}_2,\mathbf{x}_3\ldots\mathbf{x}_{2M})\Psi^*(\mathbf{x}'_1,\mathbf{x}'_2,\mathbf{x}_3\ldots\mathbf{x}_{2M})d\mathbf{x}_3\cdots d\mathbf{x}_{2M}$$

(1)

and may be written in the form

$$\rho_2(\mathbf{x}_1,\mathbf{x}_2;\mathbf{x}_1',\mathbf{x}_2') = \sum_{ij,kl} g_{ij}(\mathbf{x}_1,\mathbf{x}_2) g^*_{kl}(\mathbf{x}_1'\mathbf{x}_2') P_{ij,kl} = \mathbf{gPg}^\dagger$$

(2)

where $P_{ij,kl}$ is an element of the pair population coefficient matrix **P**. $g_{ij}(\mathbf{x}_1,\mathbf{x}_2)$ is a 2-electron Slater determinant and $\mathbf{x}_1, \mathbf{x}_2,..$ are spin-space variables normalised as in ref(26) so that the eigenvalues of the density matrix represent electron pair populations. Yang[19] showed that the existence of a superconducting condensate can be identified for a many-electron wavefunction $\Psi(\mathbf{x}_1,\mathbf{x}_2,\ldots\mathbf{x}_{2M})$ by the existence of a macroscopically large eigenvalue $\lambda_L$ of the matrix **P**. Such a macroscopically large eigenvalue of the pair density matrix corresponds to populating the same pair state with $\lambda_L$ electron pairs analogous to Bose-Einstein condensation. The eigenvector of the density matrix associated with the large eigenvalue gives the pair superconducting condensate wavefunction $\psi(\mathbf{x}_1,\mathbf{x}_2)$ as discussed more recently by Leggett[22,23].

However, in contrast to the usual BCS scenario with attractive matrix elements and perhaps counter-intuitively, a similar stabilization with the occurrence of ODLRO can occur with repulsive off-diagonal matrix elements in the Hamiltonian matrix as discussed in a paper given by us[25] over thirty years ago and subsequently developed[26,27] in which we suggested that high temperature superconductivity might arise from repulsive electronic interactions in magnetic systems from a coherent sign change in the variational coefficients. More recently, there has been wide discussion of the zero temperature BCS gap equation[28,29,13] given in k-space by

$$\Delta(\mathbf{k}) = -\sum_{\mathbf{k}'} V(\mathbf{k},\mathbf{k}') \frac{\Delta(\mathbf{k}')}{2E(\mathbf{k}')}$$

(3)

A stable superconducting state ($\Delta(\mathbf{k})$ and $E(\mathbf{k})$ are the BCS gap parameter and excitation energy) can be found for repulsive matrix elements $V(\mathbf{k},\mathbf{k}')$ if the variational coefficients $\Delta(\mathbf{k})$ change sign across the Fermi surface as might arise from anti-ferromagnetic spin fluctuations thereby allowing both sides of eqn.(3) to be positive. Although stated in different form, such a result is very closely related to eqns(6-7) of ref(25). The focus here is on how a similar and possibly equivalent result to that discussed above may be derived for the alternant cuprate and iron-based superconductors in a real space representation of the electronic wavefunction with repulsive off-diagonal matrix elements of the Hamiltonian matrix.

Following ref (25) the simplest way to understand why repulsive off-diagonal matrix elements accompanied by a sign change in the variational coefficients may lead to stable superconducting ground states is to consider the real symmetric 2$k$-dimensional Hamiltonian matrix lowest eigenvalue/eigenvector relation given in block form as

$$\mathbf{H}\begin{pmatrix} \mathbf{1} \\ -\mathbf{1} \end{pmatrix} = \begin{pmatrix} \mathbf{U} & \mathbf{V} \\ \mathbf{V} & \mathbf{U} \end{pmatrix}\begin{pmatrix} \mathbf{1} \\ -\mathbf{1} \end{pmatrix} = (U - kV)\begin{pmatrix} \mathbf{1} \\ -\mathbf{1} \end{pmatrix}$$

(4)

with the Hamiltonian matrix, $\mathbf{H}$, consisting of two blocks, $\mathbf{U}$ a $k$-dimensional diagonal matrix with diagonal elements equal to $U$, and $\mathbf{V}$ a $k$-dimensional matrix with the repulsive matrix element, $V > 0$ filling every element in the off-diagonal block $\mathbf{V}$. The vector $\pm \mathbf{1}$ is a $k$-dimensional vector with elements $\pm 1$.

The most general case is discussed in ref (25) and applies in real or momentum space. The lowest eigenvalue is $U-kV$ while the other eigenvalues are at $U+kV$ and $U$ which is $(2k-2)$ fold degenerate[26]. The lowest eigenvalue at $U-kV$ can thus cross over into a re-built ground state exactly as above for the BCS superconductor depending on the magnitudes of $U$ and $kV$. The associated ground state wavefunction also exhibits ODLRO[19,22,23,26]. No such cases of this are yet known in Nature as the conditions in which it may occur are stringent but it seems possible that both the Cuprates and Iron-based superconductors are examples of this. In the BCS case we may speak loosely of bound electron pair formation but with repulsive off-diagonal matrix elements the term 'correlated electron pairs' is more appropriate. Here the energy of the system is lowered by electrons avoiding each other at very short-range but correlating over longer distances in the attractive region of the potential to give a lower electronic energy than the normal state. We will now discuss how this may occur in iron-based and cuprate superconductors by the same mechanism in both materials.

We consider an effective electronic Hamiltonian $H$ in the random phase approximation for the electrons on the cells of a square alternant arrangement of unit cells with local point group symmetry appropriate for the superconducting layer. The long-range electronic interactions are assumed to be taken account of in the zero-point energies of the plasma modes[30] and cancelling out in any energy differences and will not be considered further. The electrons in the cores of the layer atoms and those outside the layer cannot be ignored but to include them in a rigorous fashion is prohibitively difficult. We will go some way to allow for the presence of such electrons by utilising a device widely used in semiconductor theory[31], namely that of an effective background dielectric constant $\varepsilon$ as used in ref(27).

$H$ is the sum of one-body $h(i)$ and two-body screened interactions and longer range attraction given by

$$H = \sum_i h(i) + \frac{1}{2} \sum_{i,j}{}' \left( \frac{\exp(-k_c r_{ij})}{\varepsilon r_{ij}} + f_{sr}(r_{ij}) \right)$$

(5)

$k_c$ is the inverse screening length. We will estimate $k_c$ using the inverse Thomas-Fermi screening length given by $k_c^2 = 16 \pi^2 me^2 (3 \eta_c/\pi)^{1/3} / \varepsilon h^2$ where $\eta_c$ is the carrier density and $\varepsilon$ the high frequency dielectric constant of the polarisable background taken to be about 5 as used in ref(27). The screened Coulomb interaction is supplemented by a long-range attractive tail $f_{sr}(r)$ possibly due to spin fluctuations or to Friedel oscillations[32] or some other mechanism which we do not identify. Leggett (personal communication) has suggested that a natural possible origin of the long-range attractive tail is "overscreening" of the bare Coulomb repulsion by polarisable atomic cores.

We will work in a localized Wannier type representation of the electronic basis functions where the symmetry of pair condensate wavefunctions with $d_{x^2-y^2}$ and $s\pm$ symmetries respectively in superconducting cuprates and iron-based materials has guided our choice of basis functions. In doing so we show that such choices as we discuss below, lead to low energy ground states which are driven by electron correlation and which exhibit ODLRO.

High temperature superconductors exhibit 'robust' superconductivity which is characteristic of pairing of electrons in time-reversed states as emphasized by Leggett[23]. We will be guided by this principle in our choice of pairs of Wannier functions (assumed real) to be discussed below.

## Iron Based Superconductors

Each layer is a square arrangement of $N/2$ cells with an integer index $l$ and associated with each cell are two quasi-degenerate sets labeled b1, b2 (not symmetry classifications) of symmetry adapted orthonormal localized Wannier type functions $\{\phi_{l,b1}(\mathbf{x})\}$, $\{\phi_{l,b2}(\mathbf{x})\}$ centered on each unit cell. This choice has been guided by the pair condensate wavefunction s± symmetry found in iron-based materials The expected shapes of these are shown in Fig 1(a),(b). These transform as $x^2-y^2$ and $xy$ in the near $S_4$, $C_{4v}$ and $D_{4h}$ point group symmetries[33] and resemble the xy and XY Wannier functions calculated by Andersen and Boeri[34] for pnictide superconductors derived from Fe and As multibands.

Such localized Wannier- type orbitals which in principle incorporate all crystal atomic orbitals peak on the cell on which they are centered but spread out over neighboring cells with gradually decreasing oscillations required to ensure orthogonality but may differ somewhat in vertical and lateral extensions. Each pair ($l$,b1↑, $l$,b1↓) of Wannier orbitals is assigned a signature $(-1)^l$ and each pair ($l$,b2↑, $l$,b2↓) a signature $(-1)^{l+1}$ as shown in the left panel of Fig. 2 below where positive and negative signatures are indicated. The Wannier pair functions thus labeled show an alternant pattern where each pair of Wannier functions has nearest and next-nearest neighbours with opposite signatures.

Consider a basis which describes $M$ singlet coupled electron pairs randomly distributed over $N$ Wannier orbitals where each pair is either occupied by 2 electrons or empty. The fractional filling is $\rho = M/N$.

The number of configurations of $M$ electron pairs over $N$ pairwise occupied orbitals is $\dfrac{N!}{M!(N-M)!}$.

The many electron wavefunction $\Psi(\mathbf{x_1},\mathbf{x_2},\ldots)$ may be expanded in a basis of Slater determinants $\{\phi_k\}$ where $\Psi(\mathbf{x_1},\mathbf{x_2},\ldots) = \Sigma c_k \phi_k$ where $\{c_k\}$ are the set of expansion coefficients obtained as an eigenvector of the Hamiltonian matrix. Each Slater determinant describing the configuration k of the pairwise filled Wannier orbitals on the lattice has an overall signature given by the product of all the signatures $\sigma_i$ of the occupied pairs of orbitals in the configuration (see Fig 2) such that $c_k = \prod_{i \in k} \sigma_i$.

Any one randomized configuration will have on average $5M(1-\rho)$ nearest or next-nearest neighbour interactions with basis functions of opposite signature. These will not all be equal but for simplicity all intercell pair transfer matrix elements will be given the value $V$ and intracell matrix elements $v$. The most significant interaction between Slater determinant basis functions occurs in the off-diagonal blocks of the Hamiltonian matrix. Matrix elements between two different Slater determinants basis functions are only finite when there is one Wannier pair difference in the occupation numbers. Nearest neighbour transfers and intracell pair transfers are most significant energetically and these are between configurations having opposite signatures. In such a case the off-diagonal Hamiltonian matrix element is a two-electron integral of the type

$$V = \langle \phi_{i,b1}(\mathbf{r_1}) \phi_{i,b1}(\mathbf{r_2}) | \frac{\exp(-k_c r_{12})}{\varepsilon r_{12}} + f_{sr}(r_{12}) | \phi_{j,b1}(\mathbf{r_1}) \phi_{j,b1}(\mathbf{r_2}) \rangle$$

(6)

for pair transfers between b1-b1 orbitals, and similarly for b2-b2 for different cells i,j. Intracell b1-b2 pair transfers have a corresponding matrix element

$$v = \langle \phi_{i,b1}(\mathbf{r}_1) \phi_{i,b1}(\mathbf{r}_2) | \frac{\exp(-k_c r_{12})}{\varepsilon r_{12}} + f_{sr}(r_{12}) | \phi_{i,b2}(\mathbf{r}_1) \phi_{i,b2}(\mathbf{r}_2) \rangle$$

(7)

The diagonal elements $u$ which represents the Coulomb energy to bring an electron pair into the same pair of Wannier orbitals are of the form

$$u = \langle \phi_{i,b1}(\mathbf{r}_1) \phi_{i,b1}(\mathbf{r}_2) | \frac{\exp(-k_c r_{12})}{\varepsilon r_{12}} + f_{sr}(r_{12}) | \phi_{i,b1}(\mathbf{r}_1) \phi_{i,b1}(\mathbf{r}_2) \rangle$$

(8)

and similarly for b2 cells. Matrix elements $V$ and $v$ are non-zero only by virtue of some region of orbital overlap. Hence all integrals beyond nearest neighbour interactions are neglible in a localised basis. Futhurmore, nearest neighbor b1-b1, and b2-b2 pair transfer matrix elements are by far the most significant for different cells. The contribution of $f_{sr}(r)$ to the off-diagonal matrix elements $v$ and $V$ is likely to be small if the minimum in $f_{sr}(r)$ falls outside the overlap region and hence does not contribute to the integral. However, $f_{sr}(r)$ can be very significant in lowering $u$.

It is of some importance to note that expansion of the exponential in (6-8) gives to first order for Wannier pairs $w$ and $x$

$$\langle ww | \frac{\exp(-k_c r_{12})}{\varepsilon r_{12}} | xx \rangle \approx \langle ww | \frac{1}{\varepsilon r_{12}} - \frac{k_c}{\varepsilon} | xx \rangle \quad (9)$$

provided the overlap region is small enough to give convergence of the series expansion. Using the orthogonality of the basis functions shows that $v$ and $V$ are not screened to first order but $u$ can be strongly screened. The lowest eigenvalue from such a basis is therefore expected to be close to $U - M(4V + v)(1 - \rho)$ where $U$ is the diagonal element.

In second quantized form the ground state wavefunction is given as the lowest energy eigenvector of the Hamiltonian matrix eqn(4) by

$$\Psi(\mathbf{x1}, \mathbf{x2}, \ldots) = (\sum_l (-1)^l (a^\dagger_{lb1\uparrow} a^\dagger_{lb1\downarrow} - a^\dagger_{lb2\uparrow} a^\dagger_{lb2\downarrow}))^M |0\rangle$$

(10)

We suppose that the re-built ground state competes with a normal ground state whose wavefunction is a single Slater determinant $\Psi_{normal}$. In a localized basis with screened but locally strong Coulomb repulsions interactions and maximally unpaired electrons in the normal phase, the one body energies cancel and long-range interactions cancel and an estimate of the energy difference/orbital between the normal and superconducting phases is given by

$$(1-\rho)u - (4V+v)\rho(1-\rho) \qquad \text{if } \rho > 1/2$$

$$\rho u - (4V+v)\rho(1-\rho) \qquad \text{if } \rho < 1/2$$

(11)

These expressions are symmetrical about the half filling point ($\rho = 0.5$). We expect $u$ to be strongly screened but not *(4V+ v)*. This suggests that the expressions in (11) can become negative with increased screening and thus allowing the superconducting state to become the ground state. The energy per condensed pair is shown in Fig.3 where we have taken

$$u \approx \frac{e^2}{\varepsilon L/2} \exp(-k_c L/2) , \quad (4V+v) = 0.6 \text{eV}$$

where L is the unit cell length taken to be about 4Å. The trend of appearance then disappearance of superconductivity with doping found experimentally is thus reproduced. The magnitude of this binding energy is equivalent to an energy gap is about 0.015 eV suggesting a transition temperature at optimal doping of roughly 100-200 K. Some screening of (4*V*+*v*) at higher doping will cause the curve in Fig.3 to shift to the left somewhat.

## Cuprate Superconductors

The alternant cuprate lattice has a very well known layer structure with local $C_{4v}$ symmetry as shown in Fig. 2 for a square arrangement of unit cells. In the cuprates layers, while there is no agreement as to the pairing of electrons derived from oxygen (2p) or copper (3d) bands or some mixture of these, there is on the other hand a widely held view that the pair condensate wavefunction has a dominant singlet d $_{x^2-y^2}$ symmetry ($B_1$ symmetry in the $C_{4v}$ point group). Previously, we gave a group theoretical analysis[35] of the pair condensate wavefunction in real space for cuprate superconductors with local $C_{4v}$ point group symmetry about the Cu atoms. A dominant pair of Wannier-type localized orbitals which we label as (*px,py*) with e-symmetry in the $C_{4v}$ point group, which seem very likely to be involved in the superconductivity of cuprates was identified.

To aid the reader we summarize the group theoretical analysis of the cuprate condensate wavefunction given previously[35] to which the reader is referred. The range over which the condensate wavefunction $\psi(x_1,x_2)$ remains finite is a measure of the pair size or superconducting coherence length, which is known to be only a few Ångstroms in cuprate superconductors[22,23] and such a feature may indicate some type of real space pairing. It is widely thought that in cuprate layers the Cu d-orbitals mix with oxygen $2p_x$, $2p_y$, $2p_z$ orbitals surrounding the Cu atom and that these orbitals play a dominant role in the superconductivity. We therefore considered a set of symmetry adapted orthogonalised localized basis functions $\{\phi_i(\mathbf{r})\}$ formed principally from a linear combination of atomic orbitals centered on the Cu atoms in the cuprate layer. Each localized orbital $\phi_i(\mathbf{r})$ or a degenerate pair of these centered on the same site, must transform as one of the irreducible representations (IRs) $a_1$, $a_2$, $b_1$, $b_2$, $e$ (for a degenerate pair) of the $C_{4v}$ point group.

Since the coherence length is short, it is reasonable to assume that the active orbitals $\{\phi_i(\mathbf{r})\}$ are strongly localized in space so that $\psi(x_1,x_2)$ decays to zero as $|\mathbf{r}_1 - \mathbf{r}_2|$ increases. $\psi(x_1,x_2)$ may then be decomposed into the form

$$\psi(x_1,x_2) = \chi(\Gamma) + \sum_k \varphi_k \qquad (12)$$

where $\chi(\Gamma)$ represents the pair functions centered on the principal axis transforming in the $C_{4v}$ group as the irreducible representation $\Gamma$ while the terms in the sum are a linear combination of the remaining terms transforming together as the same irreducible representation $\Gamma$.

If the condensate wavefunction $\psi(x_1,x_2)$ has $B_1$ symmetry then both terms in the right hand side of eqn.(12) above must have $B_1$ symmetry under the operations of the $C_{4v}$ group.

By reference to a direct product table for the $C_{4v}$ group[36] we can ask what pairs of irreducible representations make up $\chi(\Gamma)$ and have an antisymmetrised direct product with a $B_1$ component?

The possibilities are

$$a_1 \otimes b_1 = B_1, \qquad a_2 \otimes b_2 = B_1, \qquad e \otimes e = A_1 \oplus A_2 \oplus B_1 \oplus B_2 \qquad (13)$$

Thus, we may have (a) a degenerate pair of active e - orbitals and (b) two possible pairs from two different orbitals $a_1, b_1$ and $a_2, b_2$. The $a_1, b_1$ and $a_2, b_2$ pairs may be combined as singlets or triplets but only the singlet pairs are relevant here since experimentally the condensate wavefunction in cuprate superconductors has singlet symmetry as shown from tunneling and Knight shift studies.

The e-representation requires us to consider a degenerate pair of localized spin orbitals centered on a lattice point which we shall label *(px,py)* because of their transformation properties under the operation of the $C_{4v}$ point group. If we introduce two electrons into these orbitals we obtain four pair functions which may be combined as follows

$$
\begin{aligned}
(px(1)py(2) - py(1)px(2)) &\quad - {}^3A_2 \\
(px(1)px(2) + py(1)py(2)) &\quad - {}^1A_1 \\
(px(1)py(2) + py(1)px(2)) &\quad - {}^1B_2 \\
(px(1)px(2) - py(1)py(2)) &\quad - {}^1B_1 \qquad (14)
\end{aligned}
$$

The pair symmetries in the $C_{4v}$ group are given on the right above and only the last pair function has ${}^1B_1$ symmetry. Hence we are restricted to three possible singlet pair functions of

$^1B_1$ symmetry of which one is likely to make a dominant contribution to the condensate wavefunction.

Other things being equal, for repulsive electronic correlations the ground state of $a_1 b_1$ pairs is likely to be the triplet state, $^3B_1$, by Hund's rule. This principle energetically favours the most antisymmetrical space function and symmetrical spin function for a pair by minimising the short range Coulomb repulsion between electrons and is found to be widely applicable in molecular systems. On the other hand, with attractive forces, singlet $a_1^2$ or $b_1^2$, whichever is lower in energy, with $^1A_1$ symmetry is more favoured than $a_1 b_1$. Thus, $a_1 b_1$ pairs prepared in a singlet state would be unstable to the formation of $^3B_1$ or $^1A_1$ pairs and it is hence improbable that singlet $a_1 b_1$ pairs make a dominant contribution to the condensate wavefunction in cuprate superconductors. Similar arguments can be made against the singlet $a_2, b_2$ pair. Thus, *(px,py)* pairs derived from e-orbitals seems the most likely to dominate the condensate wavefunction. These Wannier functions are symmetry adapted combinations of the Copper and oxygen 2p atomic orbitals[37]. The shape of the *(px,py)* pair of Wannier are shown in Fig. 4 a,b. These orbitals are expected to be largely out-of-phase combinations of ligand O(2p) orbitals as discussed previously and the unoccupied pair may be regarded as a pair of oxygen holes transforming as the e-representation. This choice of pairing is composed of electrons in time-reversed states if the Wannier functions are assumed real and this is thought likely to give the most robust superconductivity.

The analysis for the cuprate superconductors follows that for the iron-based superconductors by replacing the (b1,b2) pair of orbitals with a pair transforming as the e-representation in square symmetry, *mutatis mutandis.*

Following our previous argument for the iron-based materials above we derive the ground state wavefunction for the cuprates as

$$\Psi(\mathbf{x1},\mathbf{x2},\ldots) = (\sum_{l}(-1)^l (a^{\dagger}_{lpx\uparrow}a^{\dagger}_{lpx\downarrow} - a^{\dagger}_{lpy\uparrow}a^{\dagger}_{lpy\downarrow}))^M |0\rangle \tag{15}$$

Other comments made above for the iron-based materials also apply to the cuprates.

We will now make an analysis of the condensate wavefunction in these materials.

### ODLRO and Condensate Wavefunctions

The pair population density coefficient matrix **P** has a macroscopically large eigenvalue[19,22,23] and associated eigenvector given in the relationship[26]

$$\begin{pmatrix} \frac{M}{N} & -M\frac{(N-M)}{N(N-1)} & M\frac{(N-M)}{N(N-1)} & \cdots \\ -M\frac{(N-M)}{N(N-1)} & \ddots & \vdots & \vdots \\ M\frac{(N-M)}{N(N-1)} & \vdots & \frac{M}{N} & \vdots \\ \vdots & \vdots & \vdots & \ddots \end{pmatrix} \begin{pmatrix} 1 \\ -1 \\ \vdots \\ \vdots \end{pmatrix} = (M(1-\frac{M}{N})+\frac{M}{N}) \begin{pmatrix} 1 \\ -1 \\ \vdots \\ \vdots \end{pmatrix} \tag{16}$$

The large eigenvalue $\lambda_L$ is given by

$$\lambda_L = M(1-\frac{M}{N})+\frac{M}{N} \tag{17}$$

The large eigenvalue, which represents the number of condensed electron pairs, is consistent with Yang's upper bound[18,23,41] and is related to Coleman's Extreme case[38]. Note that our result is not in conflict with Zumino's theorem[39,40] as discussed by Weiner and Ortiz, since by associating each Wannier orbital in an odd-signature-cell with an appropriate phase factor, a consistent anti-symmetrised geminal power (AGP) based on an extreme geminal with positive canonical coefficients can be derived. In addition the matrix problem (16) gives an

(*N-1*)-degenerate small eigenvalue $\lambda_S = M(M-1)/N(N-1)$. It can be proved that under the conditions of the extreme state they will not contribute to the energy stabilization. The condensate density is given by $n_s$ is proportional to $\lambda_L/N = \rho(1-\rho)$. For electron doping when $\rho \to 0$ and hole doping when $\rho \to 1$ then $n_s$ is almost linear in the electron and hole densities respectively[41] as found experimentally. The iron-based superconductor condensate wavefunction is obtained from the eigenvector above as

$$\psi(\mathbf{x}_1,\mathbf{x}_2) = \left(\rho - \rho^2\right)^{1/2} \sum_l (-1)^l \, (a^\dagger_{lb1\uparrow} a^\dagger_{lb1\downarrow} - a^\dagger_{lb2\uparrow} a^\dagger_{lb2\downarrow})|0\rangle$$

(18)

The condensate wavefunction has a singlet spin symmetry. Restricted to two-dimensions, the iron-based relative wavefunction $\psi(\mathbf{r})$ is shown in Fig 1(f) with the centre-of-mass of the pair placed at the centre of a cell. The relative coordinate is $\mathbf{r} = \mathbf{r}_1 - \mathbf{r}_2$. The range over which $\psi(\mathbf{r})$ is significant is a measure of the pair coherence length which is evidently about 10 Å.

The relative wavefunction ψ (**r**) can be seen to have sign alternating s-wave symmetry in real space. In a two-dimensional **k**-space the function χ(**k**) which is the two-dimensional Fourier transform of ψ(**r**) given by $\chi(\mathbf{k}) = \int_{-\infty}^{\infty}\int_{-\infty}^{\infty} \exp(-i\mathbf{k}\cdot\mathbf{r})\psi(\mathbf{r})d\mathbf{r}$ (and scaled to facilitate visualization.

Nb: The scale in the transformed condensate wavefunctions is in units of 2/ Å) also shows a sign alternating s-wave symmetry but the detailed shape depends on the relative extensions of the orbitals. Fig.1(c) shows the Fourier transform into k-space of the relative wavefunction shown in Fig. 1(d) where the normalized xy and $x^2$-$y^2$ components have slightly different extensions. The condensate wavefunction shown in Fig 1(c) has s± symmetry . In Fig 1(f) in which the normalized xy and $x^2$-$y^2$ components are identical in extension gives a Fourier transformed condensate wavefunction with line nodes as shown in Fig. 1(e) . The condensate wavefunction shown in Fig. 1(c) has features in common with the BCS type gap functions derived for iron-based superconductors[13](see Fig.1 in this reference). In a superconductor well described by BCS theory , χ(**k**) is proportional to Δ(**k**)/ 2E(**k**). However, such is not necessarily the case here and the Fourier components of the condensate wavefunction may not be related to the excitation spectrum in such a simple way.

For cuprates, the singlet condensate wavefunction is

$$\psi(\mathbf{x}_1,\mathbf{x}_2) = \left(\rho - \rho^2\right)^{1/2} \sum_l (-1)^l \left(a^\dagger_{lp\mathrm{x}\uparrow} a^\dagger_{lp\mathrm{x}\downarrow} - a^\dagger_{lp\mathrm{y}\uparrow} a^\dagger_{lp\mathrm{y}\downarrow}\right)|0\rangle$$

(19)

The relative condensate wavefunction in real and k-space is shown in Figs. 4. In real and k-space the condensate wavefunction has d-wave symmetry.

In conclusion we have discussed a model of iron-based and cuprate superconductors where alternant lattices allow a remarkable energetic stabilization mechanism from repulsive off-diagonal Hamiltonian matrix elements leading to the occurrence of high temperature superconductivity with $d_{x^2-y^2}$ and sign alternating s-wave or s± condensate symmetries.

Since the advent of BCS theory we have become used to the notion of attractive off-diagonal matrix elements giving rise to superconductivity. Hence it is perhaps surprising that repulsive off-diagonal matrix elements also appear to be able to bring this about. The key to understanding this is in the density matrix discussed above which describes the correlation of the electronic motion. The sign alternation in the condensate wavefunctions or the density matrix given above signifies that a 'hole' develops in regions of space around each electron keeping them apart at very short-range but allowing them to reside with higher probability in the region of the long-range attractive tail.

The same mechanism allows superconductivity to occur in both iron-based and cuprate materials.

The thermal properties of the model solved for an averaged potential will be discussed elsewhere. To obtain a reliable description of the thermal properties may require quantum Monte-Carlo simulations to be carried out in the future.

**Acknowledgements**

We wish to thank Professor Anthony Leggett for helpful comments on an earlier version of the manuscript and for pointing out the possibility of an "overscreening" effect from the polarisable background.

Thanks are also due to Professor Neil Alford, Imperial College London, for helpful discussions.

As befits our backgrounds we have taken a quantum chemistry and condensed matter theory viewpoint of this problem. Professor Per Olov Löwdin who founded this journal introduced us to the importance of the eigenvalues of the second order reduced electronic density matrix for superconductivity. For those lectures and seminars in Uppsala in that cold winter of 1978 we express our gratitude.

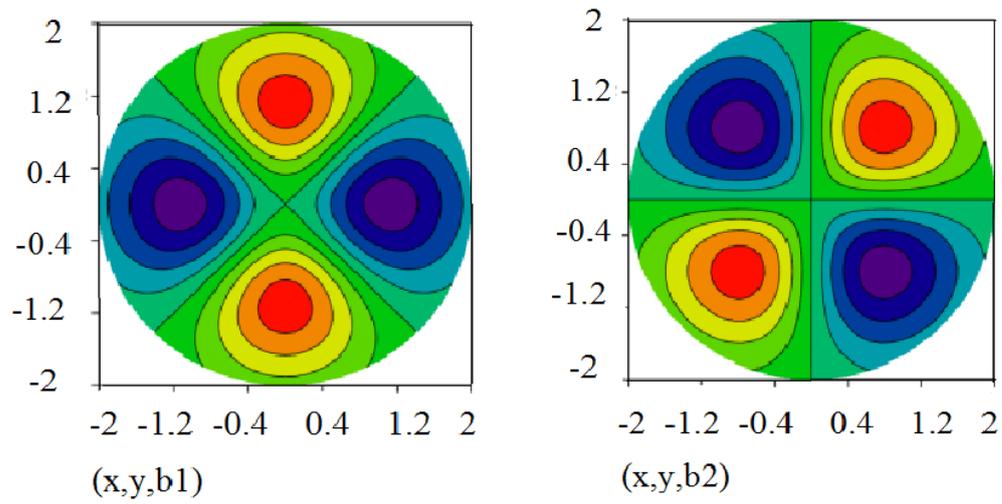

**Fig 1 (a),(b)** Shapes of b1 and b2 pairs of Wannier functions localised on each cell for iron-based superconductors. The lobes with red and blue (dark and light in greyscale) centers have opposite signs.

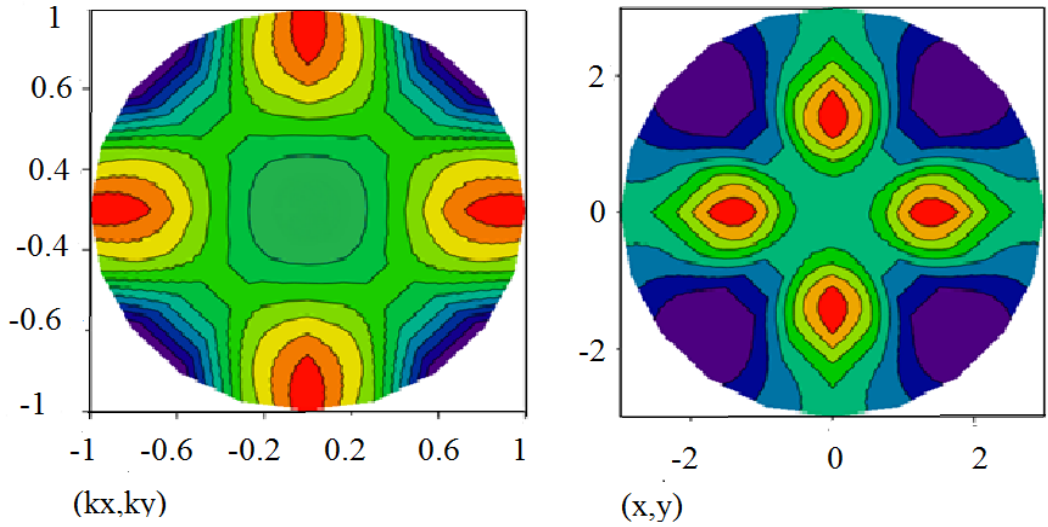

**Fig.1(c),(d)** Shapes of Iron-based Condensate wavefunction in k-space and real space . The region around the origin of the k-space condensate wavefunction (green area ,grey in greyscale), remains small and positive but is sensitive to the exact lateral extension of the lobes of the real space function where the normalized lobes in **(d)** have a slightly different extension. The lobes with red and blue (dark and light in greyscale) centers have opposite signs.

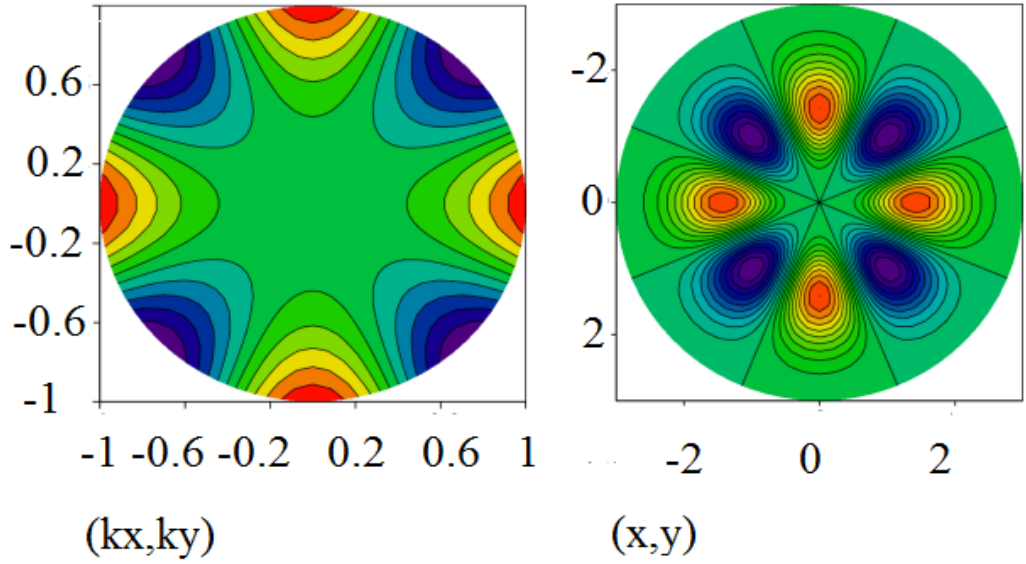

**Fig.1(e),(f)** As for 1(c),(d) but where the normalised lobes in (f) have identical extensions. The lobes with red and blue (dark and light in greyscale) centers have opposite signs. The green area (grey in greyscale) near to the origin is close to zero.

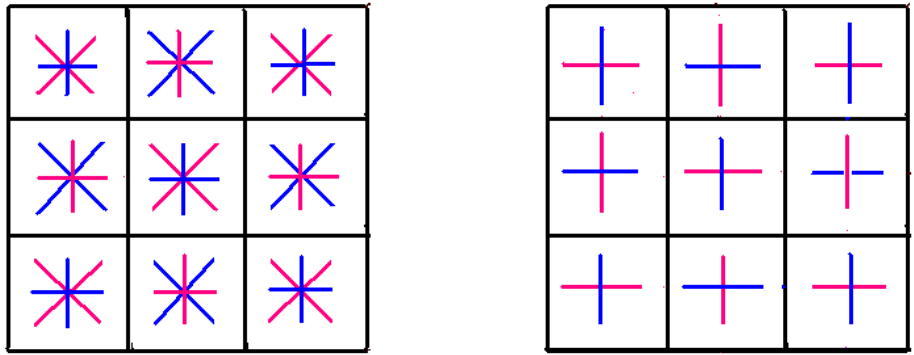

**Fig 2.** Signatures of Wannier pair functions on Iron-based (left) and cuprate (right) superconductor lattices showing nine unit cells. The Wannier pair functions are labeled with a positive (dark-greyscale, red-color online) or negative sign (light -greyscale, blue-color online). The signature of a configuration of electron pairs on the lattice is given by the product of the signs of the doubly occupied Wannier functions. Coupling matrix element are most significant between Slater determinants of opposite signature involving nearest neighbor pair transfers.

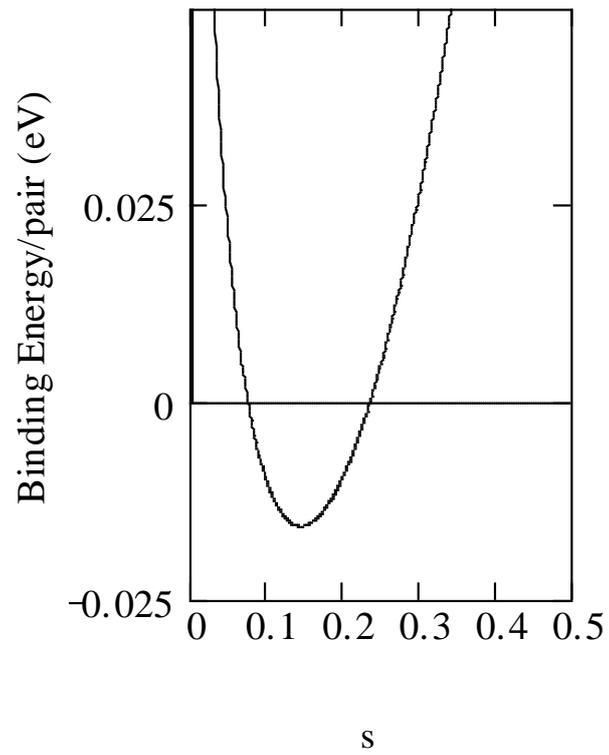

Fig 3. Binding energy /condensed pair against dopant concentration for the parameters discussed in the text. For electron doping s=ρ and for hole doping s=(1-ρ). The inclusion of some screening in (4V+v) at higher doping will shift the curve in that region to the left.

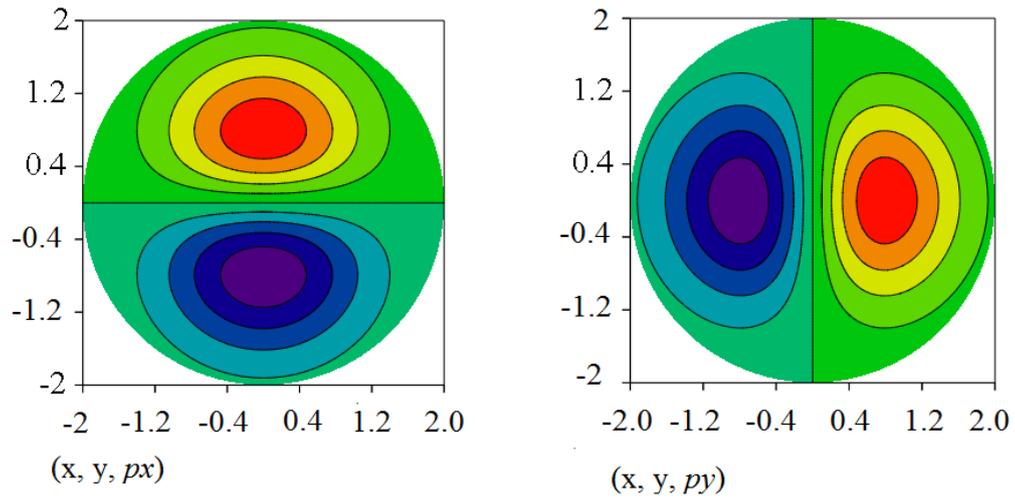

**Fig 4 (a),(b)**Shape of *px-py* pair of Wannier functions (or e-representation in square symmetry) for Cuprate superconductors. These orbitals are expected to be largely out-of-phase combinations of ligand O(2p) orbitals as discussed in the text. The lobes with red and blue (dark and light in greyscale) centers have opposite signs.

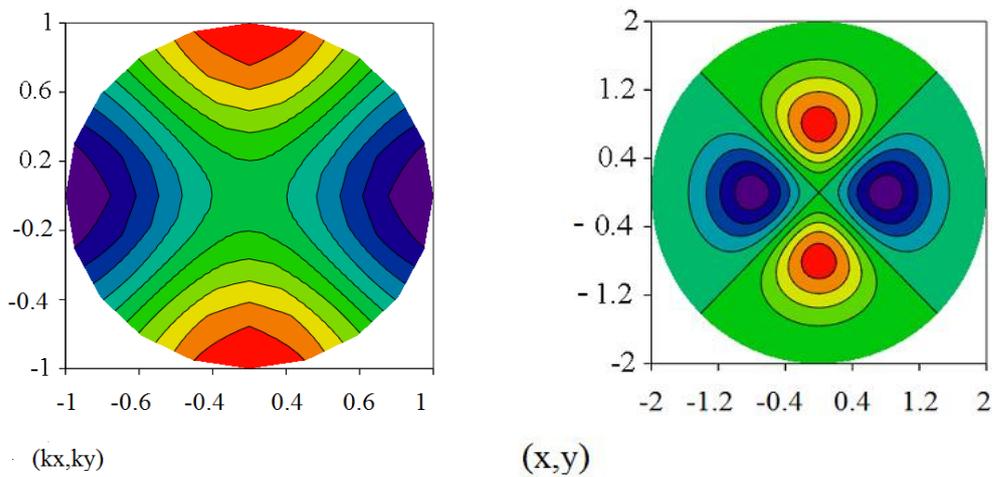

**Fig 4 (c),(d)** Shape of Cuprate condensate wavefunction in k-space and real-space. The lobes with red and blue (dark and light in greyscale) centers have opposite signs. The green areas (grey in greyscale) are close to zero.